\documentclass[aps,prd,nofootinbib,twocolumn,floatfix,
letterpaper,superscriptaddress]{revtex4}
\usepackage{graphicx}\usepackage{natbib}\usepackage{times}\usepackage{amsmath}
\usepackage[T1]{fontenc}\usepackage{textcomp}\usepackage{mathptmx}

\begin{document}

\title{Dissecting the Cygnus Region with TeV Gamma Rays and Neutrinos}

\author{John F. Beacom}
%\email{beacom@mps.ohio-state.edu}
\affiliation{Department of Physics, Ohio State University, Columbus, Ohio 43210}
\affiliation{Department of Astronomy, Ohio State University, Columbus, Ohio 43210}
\affiliation{Center for Cosmology and Astro-Particle Physics, Ohio State University, Columbus, Ohio 43210}

\author{Matthew D. Kistler}
%\email{kistler@mps.ohio-state.edu}
\affiliation{Department of Physics, Ohio State University, Columbus, Ohio 43210}
\affiliation{Center for Cosmology and Astro-Particle Physics, Ohio State University, Columbus, Ohio 43210}

\date{January 25, 2007}

\begin{abstract}
Recent Milagro observations of the Cygnus region have revealed both diffuse TeV gamma-ray emission and a bright and extended TeV source, MGRO J2019+37, which seems to lack an obvious counterpart at other wavelengths.  Additional study of this curious object also promises to provide important clues concerning one of the Milky Way's most active environments.  We point out some of the principal facts involved by following three modes of attack.  First, to gain insight into this mysterious source, we consider its relation to known objects in both the Cygnus region and the rest of the Galaxy. Second, we find that a simple hadronic model can easily accommodate Milagro's flux measurement (which is at a single energy), as well as other existing observations spanning nearly seven orders of magnitude in gamma-ray energy.  Third, since a hadronic gamma-ray spectrum necessitates an accompanying TeV neutrino flux, we show that IceCube observations may provide the first direct evidence of a Galactic cosmic-ray accelerator.
\end{abstract}

% 95.85.Ry     Neutrino, muon, pion, and other elementary particles; cosmic rays
% 98.70.Rz     gamma-ray sources; gamma-ray bursts
% 98.70.-f 	   Unidentified sources of radiation outside the Solar System
%\pacs{95.85.Ry, 98.70.Rz, 98.70.-f}
\maketitle

%--------------------------------------------------------------------%
\section{Introduction}
\begin{figure}[b!]
\includegraphics[width=3.25in,clip=true]{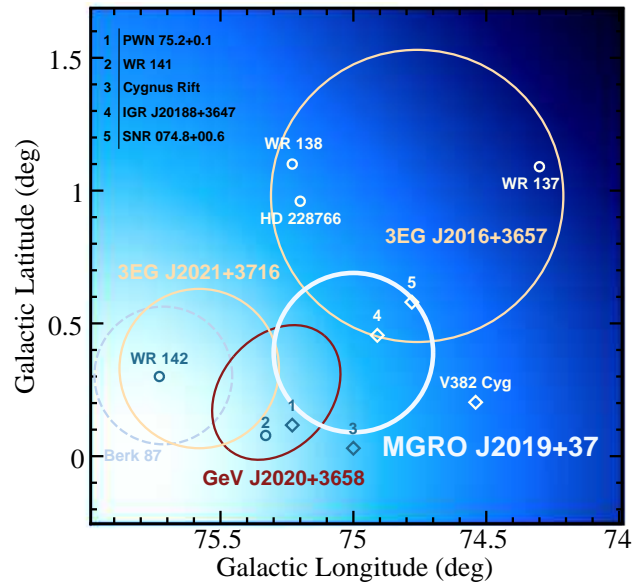}
\caption{The field near MGRO J2019+37.  Shown are 3EG Catalog sources, a GeV Catalog source, and potential counterparts to the gamma-ray sources.  These sources are overlaid upon diffuse GeV emission observed by EGRET (white most intense).  For details, please see the caption of Fig.~\ref{cyg1}.
\label{cyg2}}
\end{figure}

\begin{figure*}[t]
\includegraphics[width=7in,clip=true]{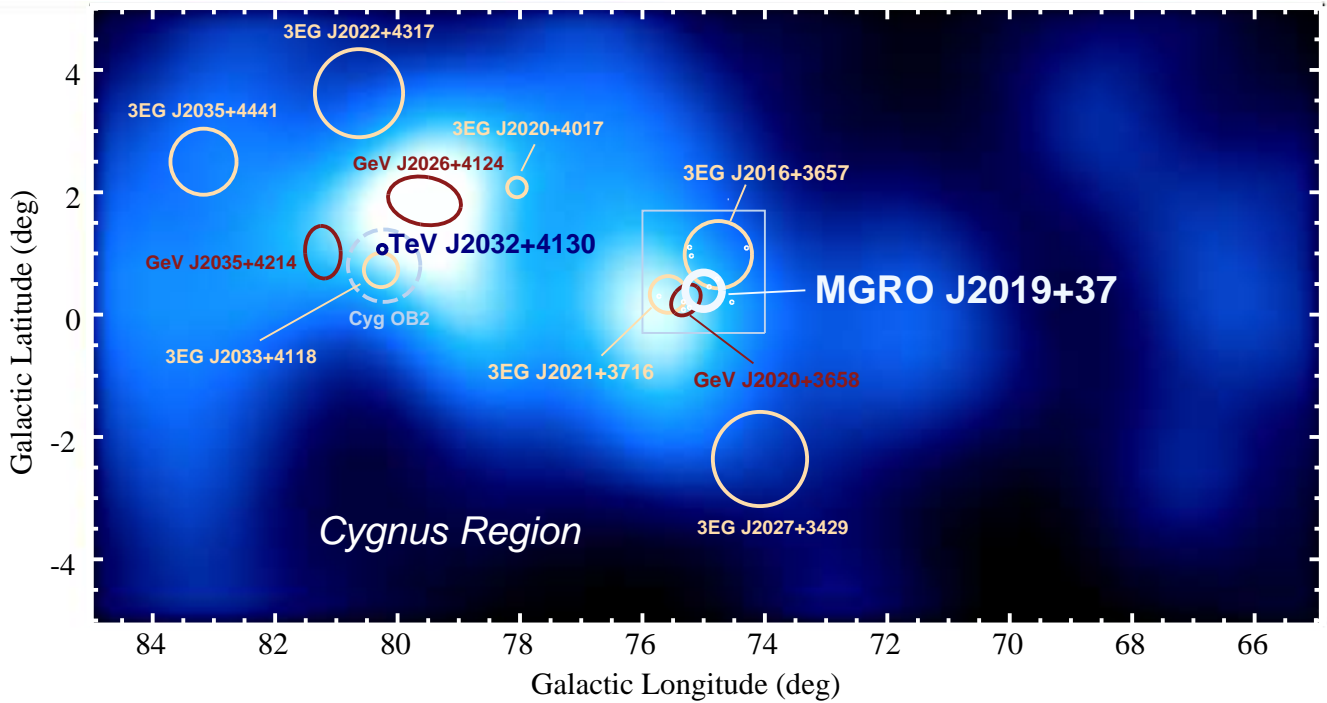}
\caption{Gamma-ray sources and diffuse GeV emission in the Cygnus region.  Shown are the sources discovered by Milagro (MGRO J2019+37) and HEGRA (TeV J2032+4130), along with their approximate (1$\sigma$) error circles.  The fitted extent of the Milagro source is comparable to the circle shown.  Also shown are nearby Third EGRET (3EG) (compiled from $> 100$~MeV gamma rays) and GeV ($> 1$~GeV gamma rays) catalog sources (all at 95\% confidence); as well as gamma-ray source candidates (points), the Cyg OB2 core (dashed circle), and the region of Fig.~\ref{cyg2} (box).  EGRET $4-10$~GeV (point-source subtracted) diffuse emission (smoothed and scaled linearly from $\sim 1-10 \times 10^{-6}$ cm$^{-2}$ s$^{-1}$ sr$^{-1}$, with white most intense) is also displayed.
\label{cyg1}}
\end{figure*}

The Cygnus region is one of the most prominent features of nearly every Galactic skymap, across many orders of magnitude in energy.  In gamma rays, sources have been known to populate this region since the time of COS~B~\cite{Hermsen:1977jj}.  In subsequent years, EGRET discovered both diffuse~\cite{Hunter:1997we} and point-like MeV-GeV emission~\cite{Hartman:1999fc}, followed by the HEGRA observation of an unidentified TeV gamma-ray source~\cite{Aharonian:2002ij}, TeV~J2032+4130.  Recently, at even higher gamma-ray energies, the Cygnus region has been discovered yet again by Milagro~\cite{Abdo:2006fq}.  The gamma-ray flux measured by Milagro includes a large ($\sim 15^\circ \times 10^\circ$) diffuse region and a new, unidentified TeV source, MGRO J2019+37 (see Fig.~\ref{cyg2}).  To clarify the relative positioning of these objects, Fig.~\ref{cyg1} shows the regions containing TeV~J2032+4130 and MGRO J2019+37 (which are separated by $\sim 5^\circ$), as well as sources from the Third EGRET~\cite{Hartman:1999fc} and GeV~\cite{Lamb:1997qs} catalogs (which all remain unidentified), overlaid on the diffuse emission measured by EGRET~\cite{Hunter:1997we, EGRET}.
%
%%%%%%%%%%%%%%%%%%%%%%%%%
%%%%%%%%%%%%%%%%%%%%%%%%%

For most TeV sources, particularly those unidentified at other wavelengths, it remains a mystery as to whether the observed gamma-ray flux is produced by hadronic or leptonic processes.  Fortunately, \textit{neutrino telescopes} offer hope for discriminating between these scenarios.  If the gamma rays arise from the the decay of neutral pions ($\pi^0 \rightarrow \gamma\gamma$) produced in proton-proton scattering, it is well-established that a corresponding flux of neutrinos (produced in $\pi^\pm$ decays) must also be present~\cite{Stecker:1978ah}.  In contrast, no neutrinos result from the inverse Compton scattering ($e^-\gamma~\rightarrow~\gamma\,e^-$) of energetic electrons on ambient photons.

HESS~\cite{Hinton:2004eu} has contributed significantly to TeV astrophysics by dramatically increasing the number of known southern-sky TeV sources (of interest~\cite{Kistler:2006hp} to a km$^3$ Mediterranean neutrino telescope~\cite{Katz:2006wv}).  Milagro's ability to survey the entire northern TeV sky is of great utility in making predictions for IceCube~\cite{Ahrens:2002dv}, which, being located at the South Pole, is well-situated to observe upgoing muons initiated by neutrinos from northern-sky sources.  The Milagro skymap~\cite{Abdo:2006fq} reveals a number of sources that will also be of interest to the ongoing and upcoming IceCube, MAGIC~\cite{Baixeras:2003xr}, VERITAS~\cite{Weekes:2001pd}, and GLAST~\cite{Gehrels:1999ri} projects, whose improved sensitivity over the previous generation of experiments will greatly enhance our understanding of Galactic cosmic-ray, gamma-ray, and neutrino production.  In particular, VERITAS is already surveying the Cygnus region in TeV gamma rays with excellent angular resolution and flux sensitivity.
%
%%%%%%%%%%%%%%%%%%%%%%%%%
%%%%%%%%%%%%%%%%%%%%%%%%%

It is of advantage to understand this region in general and MGRO J2019+37 in particular.  Towards this goal, we will first examine the vicinity of this source to note possible identified counterparts, which, among other things, give us concrete distances.  Furthermore, in addition to Milagro, this region has been observed by other experiments in the past.  We make use of these observations, particularly those of EGRET and CASA-MIA, to construct a gamma-ray spectrum for MGRO J2019+37.  Finally, we discuss the possibility of using IceCube to determine whether the observed gamma-rays are the product of hadronic processes, which would indicate a site of cosmic-ray production.
%
%--------------------------------------------------------------------%
\section{A TeV source of unknown nature}
\label{gamma}
Most conspicuous in the Milagro view of the Cygnus region is the new TeV source MGRO J2019+37 (see Fig. 3 of Ref.~\cite{Abdo:2006fq}).  While this source has been observed with very high significance ($\sim 11 \sigma$) at very high energies (median energy of 12~TeV), no obvious multi-wavelength counterpart seems to be present.  Analysis by Milagro does suggest that the TeV emission originates from either a single extended source or a combination of several unresolved point sources, fit by a 2D Gaussian of width $\theta = 0.32^\circ$.  This implies a radius of
\begin{equation}
  r \simeq 5\, \left(\frac{\theta}{0.3^\circ}\right) \,
  \left(\frac{\mathcal{D}}{1\, {\rm kpc}}\right) {\rm pc} \,,
\label{density}
\end{equation}
which depends upon the unknown source distance (scaled to 1~kpc for convenience).  We present in Fig.~\ref{cyg2} objects in the vicinity of MGRO J2019+37 that merit further study.  These objects (typically located at distances of $\sim 1-4$~kpc) can also be seen in Fig.~\ref{cyg1} in the context of the greater Cygnus region, which (for the coordinate range displayed) is also nearly completely covered by diffuse TeV emission~\cite{Abdo:2006fq}.

Of most immediate interest are two unidentified EGRET sources, 3EG J2016+3657 and 3EG J2021+3716 (constructed from gamma rays with energy $> 100$~MeV), and the GeV catalog source, GeV J2020+3658 ($> 1$~GeV gamma rays~\cite{Lamb:1997qs}).  These sources have received a fair amount of attention over the years in searches for potential counterparts.  For 3EG J2016+3657, a probable association with a blazar (of unknown redshift) was found~\cite{Mukherjee:2000ki}.  However, pair production on the extragalactic infrared background makes this an unlikely 12~TeV source~\cite{Abdo:2006fq}.  3EG J2021+3716 and GeV J2020+3658 have intriguing possible correlations with several Galactic objects.  One of these is the pulsar wind nebula, PWN 75.2+0.1~\cite{Roberts:2002qg}, which, considering the growing number of known TeV PWNe (e.g., Ref.~\cite{Aharonian:2006xx}), is also one of the better candidates for MGRO J2019+37.

In light of the recent possible discovery of a TeV source (HESS J1023-575)~\cite{Hofmann06} coincident with the Wolf-Rayet eclipsing binary WR 20a~\cite{Bonanos:2004cb}, it is worthwhile to consider similar systems near MGRO J2019+37, some of which have been examined as possible counterparts to the EGRET sources (e.g., Ref.~\cite{Romero:1999tk}).  Shown in Fig.~\ref{cyg2} are the Wolf-Rayet stars WR 137~\cite{CAMPBELL}, WR 138~\cite{CAMPBELL}, WR 141~\cite{CAMPBELL}, WR 142~\cite{Stephenson}, and HD 228766~\cite{Rauw}.  Also displayed is V382 Cyg, a massive eclipsing binary with a significant rate of mass loss~\cite{Burkholder}.  In fact, many of these systems are known to have large mass loss rates ($\gtrsim 10^{-5} M_\odot$ yr$^{-1}$~\cite{Nugis}), which may power shocks that accelerate cosmic rays.  The reader is encouraged to see Ref.~\cite{Reimer:2005hy} (and references therein) for details of gamma-ray modeling in such environments.  In a similar vein, the open star cluster Berkeley 87 has been proposed as a source of gamma rays~\cite{Polcaro}.

With the class of TeV supernova remnants already having several prominent members~\cite{Aharonian:2006ws}, such an object in Cygnus would be a compelling candidate.  However, a search of the area immediately around MGRO J2019+37 yields only SNR 074.8+00.6 (likely just an HII region~\cite{Kothes}) and the quite distant (12~kpc~\cite{Uyaniker}) SNR 074.9+01.2 (an unlikely association).  The high energy source closest to the quoted Milagro position, IGR J20188+3647, is seen only in the $17-30$~keV band~\cite{Sguera}.  Also worth noting is Cygnus Rift, a molecular cloud complex a few kpc in extent which runs east--west through this region~\cite{Uyaniker}.

%--------------------------------------------------------------------%
\section{What else is out there?}
\label{galaxy}
Thus far we have only entertained Galactic objects as potential counterparts to MGRO J2019+37.  When we examine the locations of other sources in the Milagro map of the northern sky~\cite{Abdo:2006fq}, additional support is found for this scenario.  In addition to MGRO J2019+37 and the large region of diffuse emission from Cygnus, Milagro has previously identified the Crab nebula and the blazar Markarian 421~\cite{Atkins:2003ep}, two classic TeV sources.  Milagro has also confirmed TeV emission from the vicinity of the well-studied~\cite{Aharonian:2002ij, j2032}, but poorly understood, TeV J2032+4130, which appears distinct from MGRO J2019+37.  Assuming that these sources are $\sim$~few kpc from Earth, their angular separation implies that the distance between them is at least a few hundred pc, which suggests that they are unrelated.

Since Milagro has, essentially, the same field of view as IceCube, it is also useful to note any other objects which may be of interest to neutrino (and gamma-ray) astronomy.  We consider the brightest regions in the Milagro skymap, for which identifications seem to be evident, but which were not noted in Ref.~\cite{Abdo:2006fq}.  Of course, since these regions have not yet been addressed in particular, further analysis by the Milagro Collaboration will provide the final word.  Besides the Cygnus region, along the Galactic Plane are several other regions of possible diffuse emission, which (if confirmed) would help us to further understand the ``TeV excess''~\cite{Prodanovic:2006bq} problem that arose from the initial Milagro discovery of Galactic Plane emission~\cite{Atkins:2005wu, Casanova:2006ch} and remains an outstanding concern~\cite{Abdo:2006fq}.  These are located around Galactic longitudes $\ell \approx 50^\circ$ and $\ell \approx 35^\circ$ and roughly coincide with the tangents of the Milky Way's Sagittarius and Scutum arms~\cite{Hunter:1997we, Drimmel}, respectively.  Confirmation of such emission would have tremendous value in the study of cosmic rays and their interaction with matter in the Galaxy~\cite{Steck}.

Also in the Plane, near $\ell = 40^\circ$, is a point of $> 6\sigma$ significance nearly directly coincident with GeV J1907+0557~\cite{Lamb:1997qs}.  It is intriguing that a Whipple observation of this region showed an excess of gamma-ray-like events over the entire field of view, which was deemed unlikely to be due to a bright, extended source~\cite{Fegan:2005rg}.  A number of interesting objects are located near this area, perhaps the foremost being the microquasar SS~433 and its SNR, W50, from which no TeV emission was discovered by HEGRA~\cite{ss433}.  Other systems worth noting are SNR G40.5--0.5~\cite{Sturner:1994jc}, HMXB 4U 1909+07~\cite{Seward}, candidate SNR 041.3--01.3~\cite{Gorham}, and a prominent molecular cloud complex~\cite{Huang}.  Clearly, more observations are needed in order to make any further conclusions; however, this appears to be another exciting region for future studies in high-energy astrophysics.

Since every source observed by Milagro, except for Mrk~421, is located within the Galactic Plane, these observations strongly suggest that MGRO J2019+37 is itself situated within the Milky Way.  We will proceed with further analysis following this assumption.

%--------------------------------------------------------------------%
\section{Gamma rays and cosmic rays}
\label{cosmic}
%
%%%%%%%%%%%%%%%%%%%%%%%%%
\begin{figure}[b!]
\includegraphics[width=3.25in,clip=true]{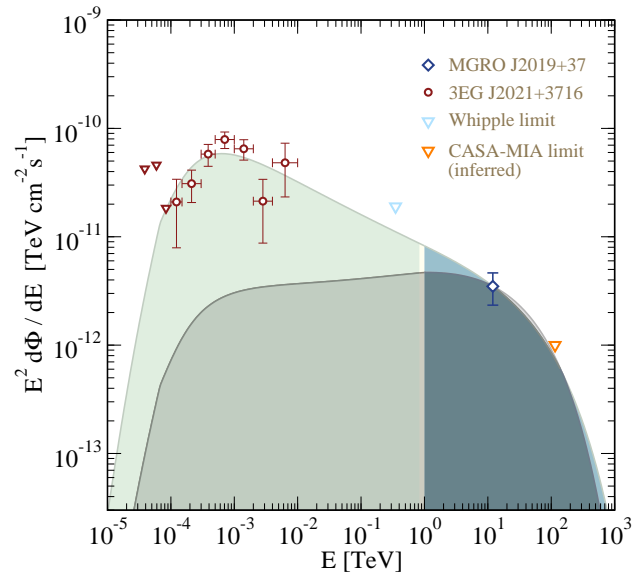}
\caption{Data and possible hadronic spectra for MGRO J2019+37.  Shown are the Milagro measurement at 12 TeV (diamond), the EGRET spectrum for 3EG J2021+3716 (circles), the upper limit from Whipple (0.3 TeV), and our inferred upper limit from CASA-MIA (115 TeV).  Also shown are hadronic fits to the data, assuming $E_p^{-2.35}$ (upper) and $E_p^{-2}$ (lower) source proton spectra.  The region above 1~TeV is of greatest interest to neutrino astronomy.
\label{cyg3}}
\end{figure}
%%%%%%%%%%%%%%%%%%%%%%%%%
%
Additional insight can be obtained by constructing a gamma-ray spectrum based on the Milagro measurement.  Milagro reported a gamma-ray flux from MGRO J2019+37 of $E^2 d\Phi/dE = (3.49 \pm 0.47_{\rm stat} \pm 1.05_{\rm sys}) \times 10^{-12}$ TeV cm$^{-2}$ s$^{-1}$ at a median energy of 12~TeV (see Fig.~\ref{cyg3}), with an undetermined (but subdominant) contribution from the surrounding diffuse emission.  This detection, at such a high energy, is quite useful in creating a spectrum, however, we also need additional data at higher and lower energies.  The two nearby (unidentified) EGRET sources may provide information at the low end.  Both have similar spectra in the MeV-GeV range~\cite{EGRET}; however, as 3EG J2021+3716 is more likely~\cite{Roberts:2000zr} to be associated with GeV J2020+3658, we will consider its possible association with the Milagro source.  Near 1~TeV, Whipple observations place upper limits on emission from the region near 3EG J2021+3716 (see Fig.~\ref{cyg3}), with a stronger limit from the vicinity of PWN 75.2+0.1~\cite{Fegan:2005rg} (not shown).  Emission from Berkeley 87 is also constrained from HEGRA observations~\cite{HEGRA} (also not shown).

At energies higher than the Milagro measurement, observational limits are crucial in determining the spectral cutoff, which is of great importance to cosmic-ray and neutrino studies~\cite{Kistler:2006hp}.  Fortunately, CASA-MIA observations can provide valuable limits.  While a previous all-sky limit is not very constraining~\cite{McKay:1993mh}, later searches had greatly enhanced statistics allowing for an improvement of more than an order of magnitude~\cite{Borione:1996jw}.  Although there is yet no published limit specifically for the MGRO J2019+37 region, we can infer a limit based upon observations of the region containing Cygnus X-3~\cite{Borione:1996jw}, as there were no significant sources in its general vicinity~\cite{Borione:1996jw, Ong}.  The integral flux limit from CASA-MIA at the position of Cygnus X-3 is $\Phi(E>115~{\rm TeV}) < 6.3 \times 10^{-15}$ photons cm$^{-2}$ s$^{-1}$, given above the median detector energy in order to reduce dependence on the assumed spectral index~\cite{Borione:1996jw}.  We infer a similar limit for the region near MGRO J2019+37 by conservatively rounding the Cygnus X-3 result up to $10^{-14}$ cm$^{-2}$ s$^{-1}$ and by treating the integral as $E\, d\Phi/dE$, giving an upper limit of $E^2 d\Phi/dE = 10^{-12}$ TeV cm$^{-2}$ s$^{-1}$ at 115~TeV for the MGRO J2019+37 region (see Fig.~\ref{cyg3}), which (even if somewhat higher) gives meaningful constraints.

As the properties of this source (e.g., magnetic fields) that would determine a leptonic spectrum are quite unknown, we will instead construct a simple, entirely hadronic spectrum (which requires fewer assumptions), not attempting to include contributions from primary or secondary electrons.  We use the parametrization for gamma rays resulting from $p$-$p$ scattering of Kelner, Aharonian and Bugayov~\cite{Kelner:2006tc}.  The gamma-ray spectrum resulting from a source distribution of protons of the form $d\Phi_p/dE_p = A_p\,E_p^{-\alpha}\,\exp\left[-\left(E_p/E_p^{\rm cut}\right)\right]$, interacting with ambient protons of density $n_H$, is 
\begin{equation}
  \frac{d\Phi_\gamma}{dE_\gamma} = c \, n_H \int_{E_\gamma}^\infty
  \sigma_{\rm inel}(E_p) \, \frac{d\Phi_p}{dE_p} \, F_\gamma\!
  \left(\frac{E_\gamma}{E_p},\,E_p\right) \frac{dE_p}{E_p}\,,
\label{spec}
\end{equation}
where $\sigma_{\rm inel}(E_p)$ is the inelastic $p$-$p$ cross section, $c$ is the speed of light, and the function $F_\gamma(E_\gamma/E_p,\,E_p)$ (given by Eq.~(58) of Ref.~\cite{Kelner:2006tc}) determines the number of photons produced per energy interval per scattering.  We use Eq.~(\ref{spec}) for $E_\gamma > 1$~TeV and the $\delta$-function approximation of Ref.~\cite{Kelner:2006tc} at lower energies to produce our gamma-ray spectra.

We first assume that there is an association between 3EG J2021+3716 and MGRO J2019+37.  Using an $E_p^{-2.35}$ proton spectrum (with $E_p^{\rm cut} = 1000$~TeV) and normalizing to the Milagro measurement at 12~TeV, we find the upper spectrum of Fig.~\ref{cyg3}.  This spectrum gives a reasonable (considering the uncertainties involved) fit to the EGRET data and, importantly, satisfies the CASA-MIA limit.  While the distance ($\mathcal{D}$) to the source and ambient proton density ($n_H$) are both unknown, normalizing to the Milagro flux allows us to find the total necessary energy injected into cosmic-ray protons as 
\begin{equation}
  \mathcal{E}_p \approx 5 \times 10^{50} \left(\frac{1\, {\rm cm}^{-3}}{n_H}\right) \,
  \left(\frac{\mathcal{D}}{1\, {\rm kpc}}\right)^2 {\rm erg} \,,
\label{energy}
\end{equation}
which, for comparison, is similar to the total explosion energy of a typical core-collapse supernova~\cite{Zwicky} when $\mathcal{D}\sim$~few kpc.

It is quite possible that neither of the EGRET sources are at all related to MGRO J2019+37.  In any case, EGRET (as well as Whipple and HEGRA) constrains the spectrum from extrapolating to lower energies too steeply, if only for the simple reason that another source would have been seen.  We also consider an $E_p^{-2}$ proton spectrum (with $E_p^{\rm cut} = 500$~TeV), which gives the lower spectrum in Fig.~\ref{cyg3}.  While safely below the low-energy observations, the CASA-MIA limit again forces a cutoff at high energies.  This scenario requires $\sim 10$~times less input cosmic-ray energy, due to the lower required GeV gamma-ray flux.

The most economical approach towards explaining the Milagro observation would be to have the gamma-ray spectrum peak (in our $E^2$ plot) at $\sim 10$~TeV and then sharply fall off (to satisfy CASA-MIA).  Such a spectrum has been observed from the PWN Vela~X by HESS~\cite{Aharonian:2006xx}, which may have a hadronic origin~\cite{Horns et al.(2006)}.  If PWN 75.2+0.1 has a spectrum of this form, it would easily satisfy the stronger Whipple limit.  However, at distance of $\sim 10$~kpc~\cite{Roberts:2002qg}, it would need to be more than an order of magnitude more energetic than Vela~X to account for the Milagro measurement.

%--------------------------------------------------------------------%
\section{Cygnus and IceCube}
\label{IceCube}
While a spectral analysis gives important guidance, the most direct way to discern the nature of a TeV source is through neutrino observations.  In high energy $p$-$p$ scattering, $\pi^+$, $\pi^-$, and $\pi^0$ are produced in roughly equal numbers~\cite{Gaisser:1990vg}.  Gamma rays result from the decay $\pi^0 \rightarrow \gamma\gamma$, while neutrinos arise from the $\pi^+\rightarrow \mu^+ \nu_\mu\rightarrow e^+ \bar{\nu}_\mu \nu_e \nu_\mu$ and $\pi^-\rightarrow \mu^- \bar{\nu}_\mu \rightarrow e^- \nu_\mu \bar{\nu}_e \bar{\nu}_\mu$ decay channels.  The ratio of neutrinos to photons is found by considering the initial neutrino flavor ratio from charged pion decay, $\nu_e:\nu_\mu:\nu_\tau = 1:2:0$, which is transformed into $1:1:1$ by vacuum neutrino oscillations before reaching Earth.  In neutrino telescopes, neutrinos and antineutrinos are practically indistinguishable, so we typically consider the sum, $\nu + \bar{\nu}$.  Each photon from $\pi^0$ decay then corresponds to one neutrino of each flavor ($N_\gamma=N_{\nu_e}=N_{\nu_\mu}=N_{\nu_\tau}$).  The typical neutrino energy is $\sim 1/2$ of the gamma-ray energy from $\pi^0$ decay.  The $\nu + \bar{\nu}$ spectrum is thus shifted, relative to a power-law gamma-ray spectrum of form $d\Phi_\gamma/dE_\gamma =
\phi_\gamma\, E_\gamma^{-\Gamma}$, as
\begin{equation}
  \frac{d\Phi_\nu}{dE_\nu} = \left(1/2\right)^{\Gamma-1}
  \phi_\gamma\,E_\nu^{-\Gamma} = \phi_\nu\,E_\nu^{-\Gamma}\,,
\end{equation}
where each neutrino flavor is treated separately.  Our interest will be limited to the $\nu_\mu + \bar{\nu}_\mu$ flux, which can produce ultrarelativistic muons in charged-current interactions within the Antarctic ice cap.

\begin{figure}[t!]
\includegraphics[width=3.25in,clip=true]{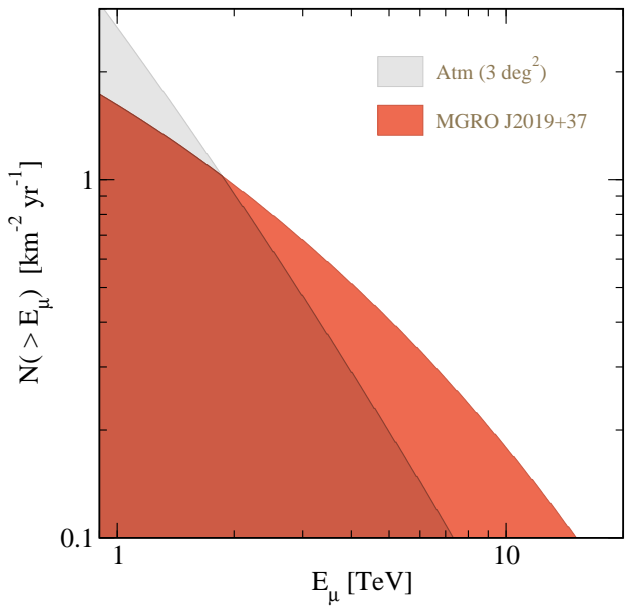}
\caption{Integrated ($\nu_\mu + \bar{\nu}_\mu$)-induced muon rates from MGRO J2019+37 above a given muon energy within IceCube for one year.  These rates result from an $E_p^{-2.35}$ source proton spectrum (see Fig.~\ref{cyg3}).  The $E_p^{-2}$ fit yields a nearly identical result.  The shaded region shows the expected atmospheric background in a 3 deg$^2$ bin.
\label{cyg4}}
\end{figure}

Knowledge of the \textit{location} of gamma-ray sources \textit{a priori} allows for neutrino searches without having to pay the trials factor associated with an undirected, all-sky search.  Knowledge of source \textit{spectra} then allows for neutrino flux predictions, which can be compared with observations.  For MGRO J2019+37, we can use the spectra shown in Fig.~\ref{cyg3} to derive the associated neutrino flux and the expected event rate of TeV muons in IceCube.  Following Ref.~\cite{Kelner:2006tc}, we parametrize each gamma-ray spectrum with a fit of the form 
\begin{equation}
  \frac{d\Phi_\gamma}{dE_\gamma} = A_\gamma \, E_\gamma^{-\beta} 
  e^{-\left(E_\gamma/E_\gamma^{\rm cut}\right)^{1/2}}\,,
\end{equation}
with $A_\gamma$ chosen to fit the gamma-ray spectrum at 1~TeV.  For the $E_p^{-2.35}$ proton spectrum, we use $\beta = 2.2$ and $E_\gamma^{\rm cut} = 45$~TeV; while we find $\beta = 1.9$ and $E_\gamma^{\rm cut} = 20$~TeV well fits the gamma rays resulting from the $E_p^{-2}$ input spectrum (above 1~TeV).  Using the methods detailed in Ref.~\cite{Kistler:2006hp}, we then calculate the expected spectrum of ($\nu_\mu + \bar{\nu}_\mu$)-induced muons (as a function of the \textit{observed} muon energy entering or originating in the detector) for IceCube, as seen in Fig.~\ref{cyg4}.  The uncertainties in these results are at or below that of the Milagro flux measurement itself.  This result is compared to the background arising from the atmospheric neutrino spectrum (which falls off more steeply with energy) in a 3~deg$^2$ bin, which roughly corresponds to the angular resolution of IceCube.  Note that we only consider muons with $E_\mu > 1$~TeV, which are more likely to have an accurately measured energy and direction than less energetic muons.  This is important for effectively discriminating between the expected signal and background.
%
%%%%%%%%%%%%%%%%%%%%%%%%%
%%%%%%%%%%%%%%%%%%%%%%%%%

With such a similar gamma-ray spectrum above 1~TeV, the $E_p^{-2}$ result is practically indistinguishable from the $E_p^{-2.35}$ case displayed.  In fact, since the Milagro measurement is at such a high energy, the expected muon rate is only weakly dependent on the spectral index, with the location of the cutoff being somewhat more important.  This rate satisfies the constraints previously derived in Ref.~\cite{Kistler:2006hp} for the Cygnus region and is consistent with the results of Ref.~\cite{Anchordoqui:2006pb}, which we recommend the reader to examine for an alternative approach.  We find that IceCube can detect neutrinos from the MGRO J2019+37 source, if indeed it is a cosmic-ray accelerator.  It is worth emphasizing that not a single high-energy neutrino source has yet been detected.  While the number of events will not be large, a signal could be separated from background within several years, especially taking advantage of the rapid increase in signal to background probability with increasing \textit{measured} muon energy.

%--------------------------------------------------------------------%
\section{Prospects and Conclusions}
Milagro has discovered a unique and very interesting object, MGRO J2019+37, in the Cygnus region of the Galaxy: a source bright in $\sim 12$~TeV gamma rays, apparently extended in size, and without clear counterparts in other wavelengths.  At present, both the nature and distance of this source remain unknown.  A near-term experimental approach that utilizes a variety of techniques is immediately suggested.  GLAST observations of MeV-GeV gamma rays should determine the validity of the EGRET discovery of several distinct sources in this region and eliminate any source confusion that may be present (as evidenced by the location of the GeV catalog source).  Whatever objects are actually present, their MeV-GeV spectra will be measured with greater accuracy.  This will be important in discerning whether a connection exists with the Milagro measurement at 12~TeV (e.g., the upper spectrum of Fig.~\ref{cyg3}).

Observations in the GeV-TeV range by VERITAS and MAGIC will provide a clear picture of the morphology of the emission and conclusive evidence as to whether single or multiple TeV sources are present.  Measurements of the shape (steep or shallow) of the gamma-ray spectrum will allow for improved source modeling and calculations of the associated neutrino flux.  Experiments like those of the Milagro and Tibet~\cite{Amenomori:2005pn} groups will play a significant role by prospecting for sources at the highest energies, and by mapping out diffuse emission.  Sometimes lost in discussions of whether hadronic TeV sources exist is the fact that the cosmic-ray protons observed in abundance at Earth must be produced \textit{somewhere} in the Galaxy.  Given the estimates of source and interstellar gas densities, pion-producing collisions should be expected.

To probe the nature of this object, we have pursued three lines of attack.  First, we considered in detail the relation of this object to others in the Cygnus region, as well as the rest of the Galaxy.  The Cygnus region is a site of intense, ongoing star formation that is dense in observed and candidate gamma-ray sources, as well as diffuse emission.  The source distance is still unknown, but could be $\sim 1-2$~kpc if associated with other objects in this region~\cite{Yadigaroglu:1996cf}.  Second, while the Milagro measurement is only in a single energy bin, we showed how the entire gamma-ray spectrum is constrained with existing observations.  We find that a simple hadronic model can easily fit EGRET data at lower energies, as well as the inferred CASA-MIA limit at higher energies.  Third, the hypothesis of a hadronic spectrum can be decisively tested by the accompanying neutrino flux, and we showed that IceCube should be able to observe this object in several years of operation.  The detection of muons initiated by TeV neutrinos would thus authenticate a Galactic cosmic-ray accelerator.

\newpage
%---------------------------------------------------------------------%
\textbf{Acknowledgments.---}%
We are grateful to Kris Stanek, Hasan Y{\"u}ksel, Brenda Dingus, Jordan Goodman, and especially Rene Ong for very helpful discussions and for comments on the manuscript.  
We also acknowledge use of the SIMBAD database.
JFB is supported by the National Science Foundation CAREER Grant PHY-0547102, MDK the Department of Energy grant DE-FG02-91ER40690, and both by CCAPP and OSU.

\vspace*{-0.35cm}
%---------------------------------------------------------------------%
%\textbf{References}

\end{document}